\documentclass[aps,pre,twocolumn,superscriptaddress,tight]{revtex4-1}

\usepackage{amssymb}
\usepackage{amsfonts}
\usepackage{color}
\usepackage{graphicx}
\usepackage{amsmath}
\usepackage{times}
\usepackage{bm}
\usepackage{float}
\usepackage{lipsum}
\usepackage{import}
\usepackage{enumerate}
\usepackage[colorlinks,linkcolor=blue,anchorcolor=blue,citecolor=blue]{hyperref}

\begin{document}

%\title{Learning multiple attractors with reservoir computing}
\title{Multi-functional reservoir computing}

\author{Yao Du}
\affiliation{School of Physics and Information Technology, Shaanxi Normal University, Xi'an 710062, China}

\author{Haibo Luo}
\affiliation{School of Physics and Information Technology, Shaanxi Normal University, Xi'an 710062, China}

\author{Jianmin Guo}
\affiliation{School of Physics and Information Technology, Shaanxi Normal University, Xi'an 710062, China}

\author{Jinghua Xiao}
\affiliation{School of Science, Beijing University of Posts and Telecommunications, Beijing 100876, China}

\author{Yizhen Yu}
\email{E-mail address: yzyu@snnu.edu.cn}
\affiliation{School of Physics and Information Technology, Shaanxi Normal University, Xi'an 710062, China}

\author{Xingang Wang}
\email{E-mail address: wangxg@snnu.edu.cn}
\affiliation{School of Physics and Information Technology, Shaanxi Normal University, Xi'an 710062, China}

\begin{abstract}
Whereas the power of reservoir computing (RC) in inferring chaotic systems has been well established in the literature, the studies are mostly restricted to mono-functional machines where the training and testing data are acquired from the same attractor. Here, using the strategies of attractor labeling and trajectory separation, we propose a new scheme of RC capable of learning multiple attractors generated by entirely different dynamics, namely multi-functional RC. Specifically, we demonstrate that by incorporating a label channel into the standard RC, a single machine is able to learn from data the dynamics of multiple chaotic attractors, while each attractor can be accurately retrieved by inputting just a scalar in the prediction phase. The dependence of the machine performance on the labeling and separation parameters is investigated, and it is found that the machine performance is optimized when the parameters take intermediate values. The working mechanism of multi-functional RC is analyzed by the method of functional networks in neuroscience, and it is revealed that each attractor is represented by a stable, unique functional network in the reservoir, and the optimal performance arises as a balance between the stability, complexity, and distinguishability of the functional networks.
\end{abstract}

\date{\today}
\maketitle

\section{introduction}

Model-free, data-based inference of chaotic systems by the technique of reservoir computing (RC), a special type of recurrent neural networks in machine learning, has drawn considerable attention in recent years~\cite{RC:Maass2002,RC:Jaeger,RC:Pathak2017,RC:LZ2018,RC:Fan,RC:digtwin,RC:DY2024,RC:adaptiveRCLYC}. From the point of view of dynamical systems, a reservoir computer can be regarded as a complex network of coupled nonlinear units which, driven by the input signals, generates the outputs through a readout function. Compared to other types of deep learning techniques such as convolutional neural networks (CNNs), RC contains only a single hidden layer, namely the reservoir. Except for the output matrix which is to be estimated from the data through a training process, all other settings of the machine are fixed at the construction, including the input matrix, the reservoir network, and the updating rules. Though structurally simple, RC has shown great potential in many data-oriented applications~\cite{RC:adaptiveRCLYC,RC:lukosevicius2009,RC:Tanaka2019}, e.g., speech recognition, channel equalization, robot control, and chaos prediction. In particular, it has been demonstrated that a properly trained RC is able to predict accurately the state evolution of typical chaotic systems for more than half a dozen Lyapunov times~\cite{RC:Jaeger,RC:Pathak2017,RC:LZ2018}, which is much longer than the prediction horizon of the traditional methods developed in nonlinear science. Besides predicting the short-term state evolution, RC is also able to replicate faithfully the long-term statistical properties of chaotic systems~\cite{RC:Pathak2017}, e.g., the dimension of strange attractors and the Lyapunov exponents. Besides the low-dimensional systems, the technique of RC has also been applied successfully to the prediction of spatiotemporal systems by a parallel scheme~\cite{RC:Pathak2018,RC:Parlitz2018,RC:ParallelMachinePRL2022}, holding promise in dealing with real-world complex systems, e.g., forecasting the atmospheric state of the entire globe~\cite{RC:AtmosphereForecastOtt2020}.

While initial studies of RC are focusing on mono-functional machines in which the training and testing data are acquired from the same attractor, recent attempts have begun to explore the capability of RC in inferring new dynamics not included in the training data~\cite{RC:CK2020,RC:Guo2021,KLW:2021,RC:Kim2021,RC:FHW2021,RC:multistability2022,RC:LHB2024}, namely the capability of knowledge transfer or transfer learning. Knowledge transfer is a vital issue for realizing artificial general intelligence (AGI) in machine learning, as it can significantly reduce the amount of data and the computational cost required in machine training~\cite{GenerativeLearning2024}. To realize knowledge transfer in RC, a practical approach is to introduce a new channel to the standard algorithm, with the input of the new channel being a scalar denoting the state of the training data (please see Ref.~\cite{RC:adaptiveRCLYC} and the references therein). Particularly, by incorporating a parameter-control channel into the standard RC, it has been demonstrated that the machine trained by the time series of several sampling states of a chaotic system is able to infer the dynamics of the other states not included in the training set~\cite{KLW:2021,RC:Kim2021,RC:FHW2021,RC:multistability2022,RC:LHB2024}. This technique, known as parameter-aware RC (PARC), has been applied successfully to the inference of typical chaotic systems, e.g., predicting the critical transition of system collapses based on the time series of few normal states~\cite{KLW:2021}, anticipating the critical coupling for synchronization in coupled oscillators by the information of several states in the desynchronization regime~\cite{RC:FHW2021}, and inferring the whole bifurcation diagram of a chaotic system using the information of a handful of states~\cite{RC:LHB2024}. Whereas the PARC technique shows the preliminary capability of knowledge transfer, the machine is still mono-functional, as the training and predicted states are governed by the same set of dynamical equations (though of different parameters). In view of the requirements of knowledge transfer in AGI, a question of both theoretical and practical interest is: Is it possible to learn multiple chaotic attractors governed by entirely different dynamics by a single reservoir computer, namely designing a multi-functional reservoir computer?

Inspired by the multifunctionality of biological neural networks, attempts have been made recently to the development of multi-functional RC in the community of nonlinear science and complex systems~\cite{RC:LZX2020,RC:Flynn2021,RC:DY2024}. In neuroscience, multifunctionality refers to the phenomenon where a single brain region or neural network is involved in multiple distinct functions~\cite{MF:PAG,MF:KLB}, which is essential for realizing a variety of cognitive functions, e.g., memory storage and retrieval. From the perspective of dynamical systems, multifunctionality means that the neural network is multi-stable, i.e., depending on the external inputs or initial conditions, the same network can be developed into completely different attractors~\cite{MF:NP}. Though various neural network models capable of generating coexisting attractors have been proposed for accomplishing multifunctionality, the models mainly adopt small-size networks with regular connections. As an emulator of the biological neural networks, RC serves as an ideal platform for accomplishing multifunctionality in large-scale, complex networks, thereby arousing the interest of researchers in nonlinear science and machine learning for designing multi-functional RCs~\cite{RC:LZX2020,RC:Flynn2021,RC:DY2024}. Leveraging the mechanism of invertible generalized synchronization, a new scheme of RC capable of learning multiple chaotic attractors is proposed in Ref.~\cite{RC:LZX2020}. In this scheme, the training data is a concatenation of the time series acquired from different attractors, and, to retrieve a specific attractor in the prediction phase, a cue series taken from the same attractor is utilized to drive the reservoir to the desired state. The proposed scheme is applied successfully to the learning of chaotic Lorenz and R\"{o}ssler oscillators, and it is found that each attractor is represented by a unique functional network in the reservoir, with the community structures of the two functional networks clearly different from each other. An alternative approach to learning multiple chaotic attractors by the RC technique is ``blending" the training data, as demonstrated in Ref.~\cite{RC:Flynn2021}. In the ``blending" scheme, the time series of different attractors are assigned with different weights, and the machine is trained by the weighted data. It is shown that by a proper setting of the weighting parameter, the machine can replicate the dynamics of different chaotic attractors accurately. Here, to retrieve a specific attractor in the prediction phase, the initial conditions of the reservoir are set as the final state of the reservoir in learning the same attractor. In neuropsychology, the schemes proposed in Refs.~\cite{RC:LZX2020,RC:Flynn2021} can be classified as implementing content-addressable memory (CAM)~\cite{CAM:1992}, as the retrieval of an attractor requires partial information acquired from the same attractor.

A different concept of memory addressing in the study of brain and computer systems is location-addressable memory (LAM)~\cite{Memory:RC}. In LAM, the information is stored at a specific physical or logical address in the memory system. To retrieve the information, one needs to know only the location or index where the data is stored. The functioning of LAM is analogous to the retrieval of information stored in labeled boxes, with the labels being the information index. In neuroscience, this storing process is underpinned by the modulation of neural synapses~\cite{Memory:RMS}, while the retrieving process is manifested as the neuronal collective behaviors induced by the inputs~\cite{Memory:LN,Memory:PK,Memory:JEK}, e.g., the neural firing patterns evocated by cues. Different from CAM, the labels of LAM are not part of the stored information and thus can be very simple, e.g., a number (in computer systems) or a concept (in the brain system). This feature makes LAM outstanding in simplicity and efficiency, especially in situations where running speed and resource-saving are of important concern. Inspired by the technique of PARC in machine learning and the concept of LAM in neuropsychology, we propose in the present work a labeling scheme of RC capable of learning multiple chaotic attractors. To be specific, we incorporate a label channel into the standard RC algorithm, and train the machine by the time series of multiple attractors together with the oscillator labels. To retrieve a specific attractor in the prediction phase, we drive the machine by inputting only the label associated with the attractor. We are able to demonstrate that such a scheme is capable of learning multiple oscillators governed by entirely different dynamics, i.e., working as a multi-functional reservoir computer. The dependence of machine performance on two key parameters distinguishing the attractors, namely the separation and labeling parameters, is investigated, and it is found that, counterintuitively, the best machine performance is achieved when both parameters take the intermediate values. By the method of functional networks in neuroscience~\cite{RC:LZX2020,FN:CR2005,FN:EB2009,FN:HSY2023}, we conduct an analysis on the working mechanism of multi-functional RC. It is revealed that each attractor corresponds to a stable, unique functional network in the reservoir, and, interestingly, the optimal performance appears as a competition between the stability, complexity, and distinguishability of the functional networks. (After the completion of the current study, we noticed that a similar scheme of multi-functional RC has been proposed independently~\cite{KLW2024}. The main difference between the two studies is that our present work focuses on the working mechanism of multi-functional RC, whereas Ref.~\cite{KLW2024} focuses on the capacity and controllability of the machine.)

The rest of the paper is organized as follows. The new scheme of multi-functional RC will be articulated in Sec. II. The application of the proposed learning scheme to multiple chaotic attractors will be demonstrated in Sec. III, together with the dependence of the machine performance on the two key parameters used for distinguishing the attractors. The working mechanism of multi-functional RC will be analyzed by the method of functional networks in Sec. IV. Generalization of the findings to other chaotic systems and the bifurcation behaviors of the reservoir will be discussed in Sec. V, together with the conclusion.

%\section{Multi-functional RC}
\section{method}

Similar to the technique of PARC~\cite{KLW:2021,RC:Kim2021,RC:FHW2021}, the multi-functional RC proposed for learning multiple chaotic attractors also consists of four modules: the input layer (input-to-reservoir), the label channel, the reservoir network, and the output layer (reservoir-to-output). The input layer is characterized by the matrix $\bm{W}_{in}\in\mathbb{R}^{D_r\times D_{in}}$, which couples the input vector $\bm{u}_{\beta_i}(t)\in\mathbb{R}^{D_{in}}$ into the reservoir network. Here, $\bm{u}_{\beta_i}(t)$ denotes that the input vector is acquired from the $i$th attractor, with $i=1,\ldots,m$ the attractor index and $\beta_i$ the label of attractor $i$. The elements of $\bm {W}_{in}$ are randomly drawn from a uniform distribution within the range $[-\sigma, \sigma]$. The label channel is coupled to the reservoir through the vector $\bm{s}=\beta\bm{W}_b$, with $\beta=\{\beta_i\}$ the attractor labels and $\bm{W}_b\in \mathbb{R}^{D_{r}}$ the bias vector. The elements of $\bm{W}_b$ are also drawn randomly from the range $[-\sigma, \sigma]$. The reservoir network is composed of $D_r$ nodes, with the initial states of the nodes being randomly chosen from the interval $[-1,1]$. The state of the reservoir network, $\bm{r}(t)\in \mathbb{R}^{D_r}$, is updated as
\begin{equation}\label{rc1}
\bm{r}(t+\Delta t)=(1-\alpha)\bm{r}(t)+\alpha\tanh[\bm {A}\bm{r}(t)+\bm{W}_{in}\bm{u}_{\beta}(t)+\beta\bm{W}_b].
\end{equation}
Here, $\Delta t$ is the time step for updating the reservoir, $\alpha\in (0,1]$ is the leaking rate, $\bm{A}\in \mathbb{R}^{D_r\times D_r}$ is a weighted matrix capturing the coupling relationship between nodes inside the reservoir. The matrix $\bm{A}$ is constructed as a sparse random Erd\"{o}s-R\'{e}nyi network: with the probability $p$, each element of the matrix is arranged a nonzero value drawn randomly from the interval $[-1,1]$. The matrix $\bm{A}$ is rescaled to make its spectral radius equal $\lambda$. The output layer is characterized by the matrix $\bm{W}_{out}\in \mathbb{R}^{D_{out}\times D_{r}}$, which generates the output vector, $\bm{v}(t)\in \mathbb{R}^{D_{out}}$, according to the equation
\begin{equation}\label{rc2}
\bm{v}(t+\Delta t)=\bm{W}_{out}\bm{\tilde{r}}(t+\Delta t),
\end{equation}
with $\bm{\tilde{r}}\in \mathbb{R}^{D_r}$ the new state vector constructed from the reservoir state (i.e., $\tilde{r}_i=r_i$ for the odd nodes and $\tilde{r}_i=r_i^2$ for the even nodes~\cite{RC:Pathak2018}), and $\bm{W}_{out}$ the output matrix to be estimated through the training process. Except $\bm{W}_{out}$, all other parameters of the machine, e.g., $\bm{W}_{in}$, $\bm{A}$ and $\bm{W}_b$, are fixed at the construction. For the sake of simplicity, we set in the current study $D_{out}=D_{in}$~\cite{RC:Pathak2017,RC:LZ2018}.

The implementation of multi-functional RC includes three phases: training, validation, and retrieval. The mission of the training phase is to find a suitable output matrix $\bm{W}_{out}$ so that the output vector $\bm{v}(t+\Delta t)$ as calculated by Eq.~(\ref{rc2}) is as close as possible to the input vector $\bm{u}(t+\Delta t)$ for $t=(\tau+1)\Delta t,\ldots,(\tau+L)\Delta t$, with $T_0=\tau\Delta t$ the transient period (to remove the impact of the initial conditions of the reservoir) and $L$ the length of the training series. This is done by minimizing the cost function with respect to $\bm{W}_{out}$, which gives~\cite{RC:LZ2018,RC:Pathak2017,RC:Fan}
\begin{equation}\label{rc3}
\bm{W}_{out}=\bm{U}\bm{V}^T(\bm{V}\bm{V}^T+\xi\mathbb{I})^{-1}.
\end{equation}
Here, $\bm{V}\in \mathbb{R}^{D_{r}\times L}$ is the state matrix whose $k$th column is $\bm{\tilde{r}}[(\tau+k)\Delta t]$, $\bm{U}\in \mathbb{R}^{D_{out}\times L}$ is another state matrix whose $k$th column is $\bm{u}[(\tau+k)\Delta t]$, $\mathbb{I}$ is the identity matrix, and $\xi$ is the ridge regression parameter for avoiding overfitting.

It is worth mentioning that the training data contains two different time series: (1) the input vector $\mathbf{u}_{\beta}(t)$ representing the state of the attractors and (2) the scalar $\beta(t)$ labeling the attractor from which $\mathbf{u}(t)$ is acquired. More specifically, the input vector $\mathbf{u}_{\beta}(t)$ is composed of $m$ segments of equal length, while each segment is a time series obtained from a specific attractor. As such, the training data is a concatenation of the times series of $m$ different attractors, and $\beta(t)$ is a step-function of time.

The machine performing well on the training data might not perform equally well on the testing data. The finding of the optimal machine performing well on both the training and testing data is the mission for the validation phase. The set of hyperparameters to be optimized in the machine include $D_{r}$ (the size of the reservoir network), $p$ (the connecting density of the reservoir network), $\sigma$ (the range defining the input matrix and the bias vector), $\lambda$ (the spectral radius of the matrix $\bm{A}$), $\xi$ (the regression coefficient), and $\alpha$ (the leaking rate). To find the optimal hyperparameters, we first set the input of the label channel to a specific value $\beta \in \{\beta_i\}$, then evolve the machine as an autonomous dynamical system by taking the output vector $\bm{v}(t)$ as the next input vector $\bm{u}_{\beta}(t)$. The $i$th attractor labeled $\beta_i$ then is reconstructed according to the output $\bm{v}(t)$. The accuracy of the retrieved attractor is evaluated by the deviation value~\cite{Noise:SR2023},
\begin{equation}\label{dv}
D\equiv\sum_{i=1}^{m_x}\sum_{j=1}^{m_y}\sqrt{(f_{i,j}-\hat{f}_{i,j})^2},\nonumber
\end{equation}
where $m_x$ and $m_y$ are, respectively, the total numbers of cells in the $x$ and $y$ directions, and $f_{i,j}$ and $\hat{f}_{i,j}$ are, respectively, the visiting frequencies of the true and retrieved trajectories to the cell $(i,j)$. [For simplicity, we consider only the trajectory projected onto the $(x,y)$-plane.] Clearly, the smaller the $D$, the more accurate the retrieved attractor. We calculate $D$ for each of the retrieved attractors ($m$ in total), and the machine's overall performance is qualified by averaging $D$ over the attractors, which is adopted as the objective function for searching the optimal hyperparameters. In simulations, the optimal hyperparameters are obtained by scanning each hyperparameter over a certain range in the parameter space using conventional optimization algorithms such as the Bayesian and surrogate optimization algorithms
~\cite{Optimization:LY}. The set of optimal hyperparameters, together with the associated output matrix, define the optimal machine, which will be used later for attractor retrieval. Finally, to retrieve attractor $i$ in the prediction phase, we set $\beta_i$ as the input of the label channel and operate the machine in the closed-form by replacing $\bm{u}_{\beta}(t)$ with $\bm{v}(t)$ in Eq.~(\ref{rc1}). The output $\bm{v}(t)$ gives the retrieved attractor. [We note that when running the machine in each phase, the reservoir network always starts from the independent and identically distributed (i.i.d.) random initial conditions. As such, the machine's performance is evaluated by the probability of successful retrieval of the attractors when the reservoir is started from random initial conditions.]

\section{Results}

\begin{figure*}[tbp]
\begin{center}
\includegraphics[width=0.95\linewidth]{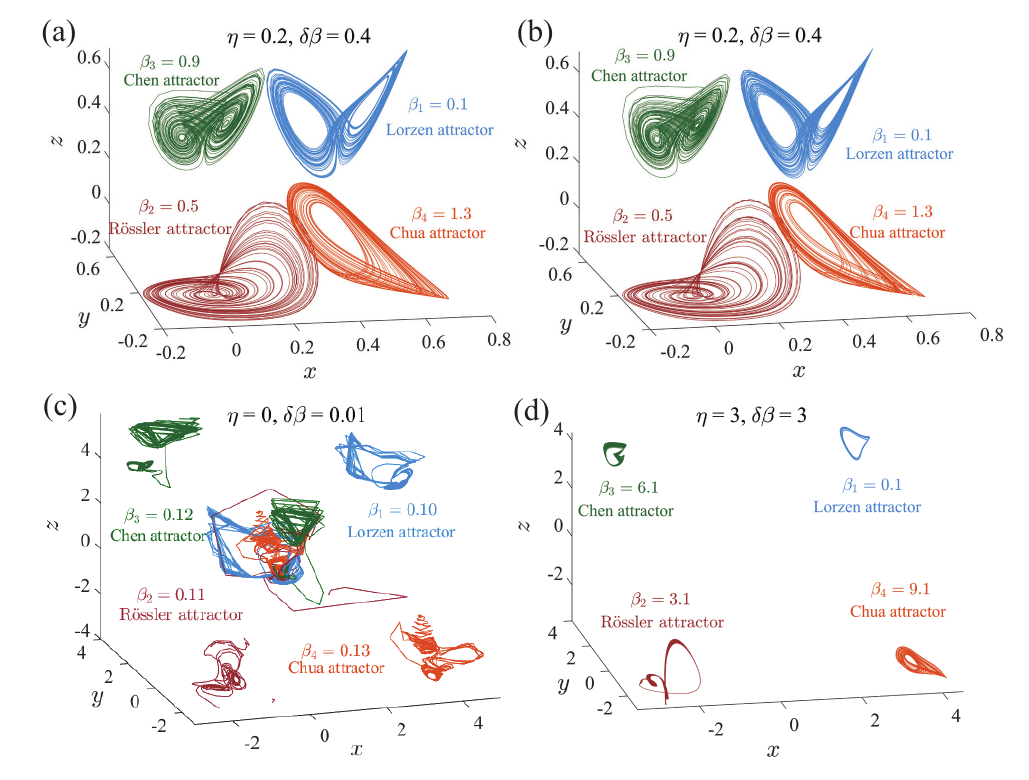}
\centering\caption{Learning multiple chaotic attractors with multi-functional RC. (a) The training attractors generated by the distinguishing parameters $(\eta,\delta\beta)=(0.2,0.4)$. $\beta_i$ is the label of attractor $i$. $\eta$ denotes the separation parameter. (b) By the parameters $(\eta,\delta\beta)=(0.2,0.4)$, all attractors are accurately retrieved from the machine. (c) By the parameters $(\eta,\delta\beta)=(0,0.01)$, the machine fails to retrieve the attractors. Plotted in the insets are the individual attractors. (d) By the parameters $(\eta,\delta\beta)=(3,3)$, the Chua attractor is accurately retrieved, while the retrievals of the other attractors are failed.}
\vspace{-0.3cm}
\label{fig1}
\end{center}
\end{figure*}

To illustrate the implementation and also check the efficacy of the multi-functional RC proposed above, we apply it to the storage and retrieval of $m=4$ chaotic attractors governed by different dynamics, including the Lorenz attractor, the R\"{o}ssler attractor, the Chen attractor, and the Chua attractor. (The models of the chaotic attractors and the details of the data processing are given in the Appendix.) Before presenting the detailed results, we first introduce two special skills adopted in processing the input data, one for separating the attractors in the phase space and the other for distinguishing the attractors in the label channel. The attractors are separated from each other in the phase space by shifting them in opposite directions. To be specific, the states $(x,y,z)$ of the Lorenz, R\"{o}ssler, Chen and Chua systems are replaced by $(x+\eta,y+\eta,z+\eta)$, $(x-\eta,y-\eta,z-\eta)$, $(x-\eta,y+\eta,z+\eta)$, and $(x+\eta,y-\eta,z-\eta)$, respectively, with $\eta$ the separation parameter. In addition, to distinguish the attractors in the label channel, we set $\beta_i=\beta_0+(i-1)\delta\beta$, with $i=1,\ldots,m$ the attractor index and $\delta\beta$ the labeling parameter. Without the loss of generality, we set $\beta_0=0.1$ in our studies and distinguish the attractors by increasing $\delta\beta$. Shown in Fig.~\ref{fig1}(a) are the attractors generated by the distinguishing parameters $\eta=0.2$ and $\delta\beta=0.4$. Clearly, the larger the parameters $\eta$ and $\delta\beta$, the clearer the attractors are distinguished from each other, which, according to the studies in neuropsychology, is helpful for the storage and retrieval of multiple memories~\cite{Memory:RC,Memory:RMS}. In what follows, we are going to demonstrate that by a proper setting of the distinguishing parameters, the dynamics of different chaotic attractors can be learned by a single machine, i.e., the attractors can be accurately retrieved by inputting just the corresponding labels in the retrieval phase, thereby accomplishing the functionality of multi-functional RC.

In preparing the datasets, we collect from each attractor a time series containing $\hat{T}=9 \times10^3$ successive states. Each time series is divided into two segments. The first segment contains $T=3 \times 10^3$ data points, which are used as the training data for calculating the output matrix $\bm{W}_{out}$. The second segment contains $T'=6\times 10^3$ data points, which are used as the validation data for optimizing the machine hyperparameters. The data are normalized (by setting the oscillating amplitude of the variables as $0.5$) and processed by the skills mentioned above (attractor shifting and labeling). The processed data are then fed into the machine to train the output matrix $\bm{W}_{out}$. The transient period for removing the impact of the initial condition of the reservoir is chosen as $\tau=200$ (the machine performance is almost unchanged by varying $\tau$, given that $\tau$ is not very small). In the validation phase, we use $\beta_i$ as the input of the label channel, and operate the machine in the closed form. After a transient period of $\tau$ steps, we collect the outputs for $T'$ steps and then calculate the deviation value, $D_i$. The value of $D_i$ is averaged over $100$ reservoir realizations (with different initial conditions), and the total number of cells used in calculating $D_i$ are $m_x=m_y=100$. This is done for each attractor, and the average of $D$ is used as the objective function for searching the optimal hyperparameters. In our studies, we fix the reservoir size as $D_{in}=100$ while optimizing only the hyperparameters $(p,\lambda,\sigma,\alpha,\xi)$. The optimal hyperparameters are obtained after $300$ trials in the parameter space with the help of the ``optimoptions" function in Matlab. The ranges over which the hyparameters are searched are: $p\in (0,1)$, $\lambda\in(0,1)$, $\sigma\in(0,3)$, $\alpha\in(0,1)$, and $\xi\in(1\times 10^{-10},1\times 10^{-2})$. The validation phase ends up with the set of optimal hyperparameters $(p,\lambda,\sigma,\alpha,\xi)=(0.66,0.39,1.77,0.66,6.4\times 10^{-3})$, which, together with the associated output matrix, defines the optimal machine of multi-functional RC.

In the retrieval phase, we input the attractor label $\beta_i$ in the control channel and operate the machine in the closed-loop configuration, with the initial conditions of the reservoir network being randomly chosen. The retrieved attractors are shown in Fig.~\ref{fig1}(b). We see that the retrieved attractors closely resemble the training attractors plotted in Fig.~\ref{fig1}(a). To evaluate the machine performance in the retrieval phase, we calculate the deviation value for $1000$ reservoir realizations, and the fraction of cases with $D$ smaller to a threshold value, $D_c$, is defined as the success rate. For the threshold $D_c=0.4$ (by which the retrieved attractors closely resemble the training attractors), numerical results show that the success rates are $100\%$ for the Lorenz, R\"{o}ssler and Chen oscillators, while it is about $89\%$ for the Chua oscillator. We thus see that for the distinguishing parameters $(\eta,\delta\beta)=(0.2,0.4)$, the machine indeed is capable of learning multiple chaotic attractors.

We proceed to investigate the dependence of the machine performance on the distinguishing parameters $(\eta,\delta\beta)$. As the two parameters quantify the distinguishability of the attractors, a natural expectation will be that the machine performance will be improved by increasing $\eta$ or $\delta\beta$. To check it out, we set $\eta=0$ (without attractor shifting) and $\delta\beta=0.01$ in processing the data, and calculate again the success rates of attractor retrievals. For this set of parameters, the success rates are all $0$. That is, the machine fails to retrieve the attractors completely. Typical cases of the failed retrievals are shown in Fig.~\ref{fig1}(c). We see that the retrieved attractors present complicated structures, but are clearly different from the training attractors. Moreover, the retrievals are sensitive to the initial conditions, and the retrieved attractor is unstable. Specifically, by changing the initial conditions of the reservoir network, the machine will output a very different attractor, and, in the long-term running, the retrieved trajectory is ``hopping" randomly between different attractors. From the viewpoint of brain memory function~\cite{Memory:RC,Memory:RMS}, the failed retrievals are understandable, as memories of similar contents (reflected as smaller values of $\eta$ and $\delta\beta$ here) are prone to confusion.

\begin{figure}[tbp]
\begin{center}
\includegraphics[width=0.8\linewidth]{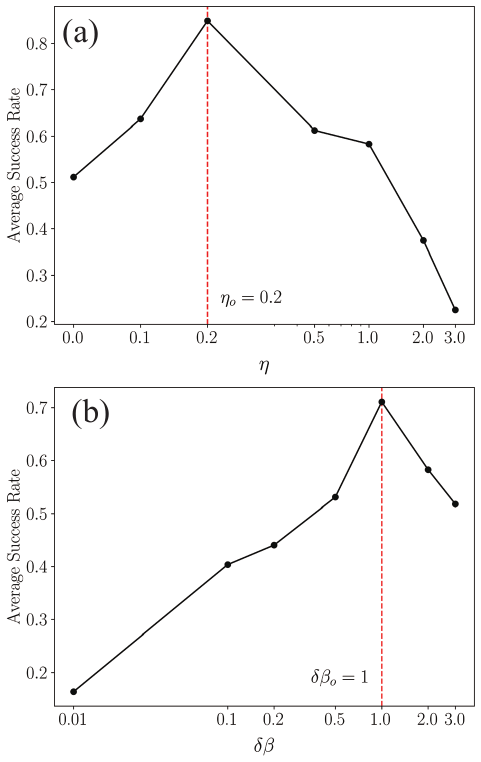}
\centering\caption{The dependence of the machine performance on the distinguishing parameters $\eta$ and $\delta\beta$. Shown are the variation of the averaged success rate of attractor retrievals with respect to (a) the attractor separation parameter $\eta$ by fixing $\delta\beta=0.4$ and (b) the attractor labeling parameter $\delta\beta$ by fixing $\eta=0.2$. Vertical dashed lines denote the optimal parameters where the averaged success rate reaches the maximum. Please note the logarithmic scales of the distinguishing parameters.}
\vspace{-0.6cm}
\label{fig2}
\end{center}
\end{figure}

To check whether the machine performance can be further improved by adopting large values of $\eta$ and $\delta\beta$, we set $\eta=\delta\beta=3$ (by which the attractors are distinctly separated from each other) and plot in Fig.~\ref{fig1}(d) the typical attractors retrieved from the machine. Counterintuitively, we see that the retrievals are deteriorated compared to the results for $(\eta,\delta\beta)=(0.2,0.4)$. A close look at Fig.~\ref{fig1}(d) shows that the retrieved attractors, while well separated from each other in the phase space, are periodic, i.e., they maintain the basic shapes of the training attractors but fail in replicating the fine structures. Different from the results shown in Fig.~\ref{fig1}(c), we find that the retrieved attractors are very stable, i.e., the attractors are independent of the initial conditions of the reservoir. Numerically, the success rates for retrieving the Lorenz and Chua attractor are $33.7\%$ and $39.3\%$, respectively, while are $0$ for the R\"{o}ssler and Chen oscillators. To have a global picture on the impact of the distinguishing parameters on the machine performance, we plot in Figs.~\ref{fig2}(a) and (b) the averaged success rate with respect to $\eta$ and $\delta\beta$. The results show that the machine performance is optimized at intermediate values of $\eta$ or $\delta\beta$. For the separation parameter, the optimal performance is achieved at about $\eta_o=0.2$ (fixing $\delta\beta=0.4$), while for the labeling parameter, the performance is optimized around $\delta\beta_o=1.0$ (fixing $\eta=0.2$). (Please note that in plotting Fig.~\ref{fig2}, a new machine is trained and optimized for each value of $\eta$ and $\delta\beta$, which involves extensive simulations.)

How can a single machine is able to learn multiple chaotic attractors, and why does the machine performance deteriorate at large distinguishing parameters? Although the preliminary results show that the machine can indeed learn multiple attractors, the internal representation of these attractors in the reservoir network and the mechanism behind optimal performance remains unclear. From an application standpoint, the distinguishing parameters could be treated as additional hyperparameters, with their optimal values determined through the established optimization algorithms. However, from the viewpoint of interpretability, it is crucial to interpret the numerical results theoretically. Motivated by this, we investigate in the following section the working mechanism of multi-functional RC through the lens of functional networks in neuroscience.

\section{Mechanism analysis}

Roughly, functional networks represent the correlation relationships among coupled dynamical units in complex systems~\cite{FN:CR2005,FN:EB2009,FN:HSY2023}. Unlike structural networks where links represent physical connections, links in functional networks are virtual and dependent on the definition of correlations. Depending on the definition of unit correlations and the criteria for establishing links, different functional networks can be derived from the same structural network~\cite{FN:ZCS2006,FN:LMH2010,FN:LWJ2015}. Functional networks have proven to be efficient for detecting spatiotemporal behaviors emerged in high-dimensional complex systems, and have been widely used in the literature for exploring the functionality of various real-world systems, such as cognitive functions of the human brain~\cite{FN:BB1995,FN:KJF2011,FN:EVM2005}. Moreover, recent advances in network science have introduced a suite of sophisticated tools for analyzing complex networks, further solidifying the functional network as a powerful framework for investigating the functions and performance of networked systems~\cite{CN:Newman,CN:BS}.

As an emulator of biological neural networks, reservoir computer shares many key features with neural systems, making functional networks also a powerful tool for investigating its working mechanism. Treating the inputs as external stimuli and the reservoir as a dynamical complex network, a necessary condition for the machine to be able to replicate the target system accurately is the establishment of generalized synchronization between the stimuli and the reservoir~\cite{GS:Rulkov1995,RC:GSTL2019,RC:LZX2020,RC:VerzelliChaos2021}. In this scenario, the long-term responses of the reservoir network are uniquely determined by the inputs, with each time series represented by a distinct functional network inside the reservoir. Leveraging this concept, recent studies have provided important insights into the working mechanisms of reservoir computers. For example, it is shown that when the machine is adequately trained and is able to replicate the target system accurately, criticality can emerge in the reservoir, i.e., the sizes of the functional network motifs follow a power-law distribution~\cite{RC:WL2022}. In Ref.~\cite{RC:LZX2020}, it is demonstrated that functional networks representing different chaotic systems are clearly different from one another, highlighting the distinct responses of the reservoir to different inputs. Inspired by these studies, we employ functional networks to analyze the working mechanisms of the new algorithm of multi-functional RC, focusing on the following two questions: (1) How are different attractors represented inside the reservoir? (2) Why is the machine performance optimized for intermediate distinguishing parameters?

\begin{figure*}[bpt]
\begin{center}
\includegraphics[width=0.98\linewidth]{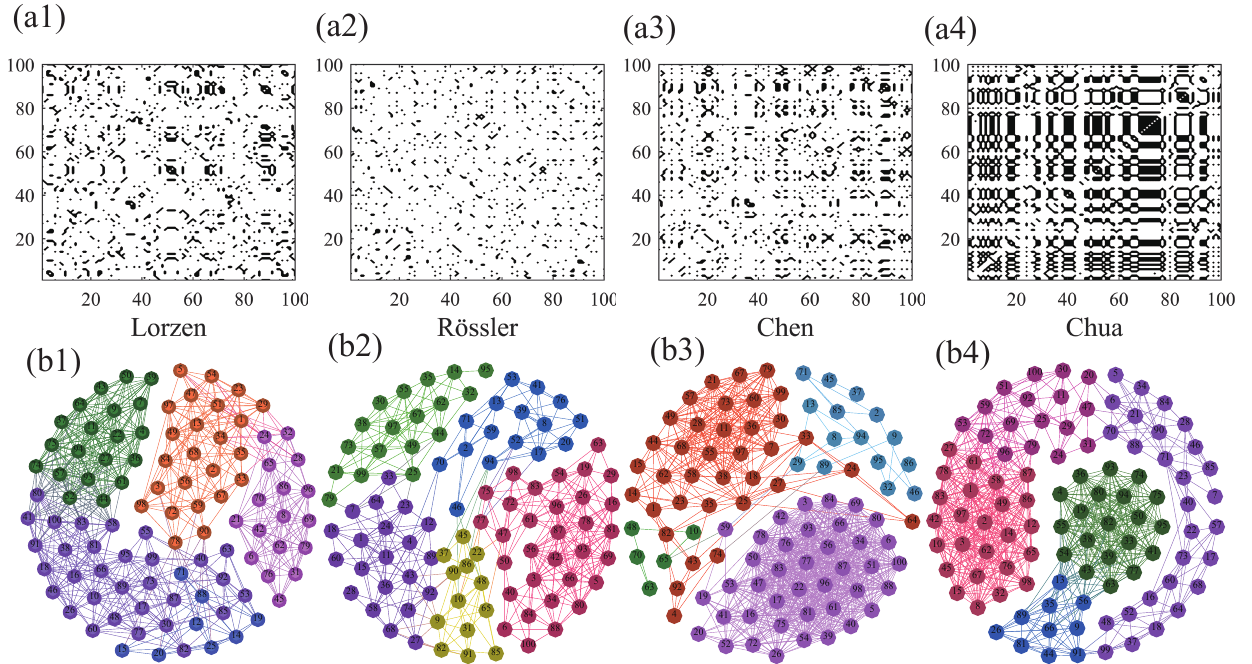}
\caption{Functional networks obtained in retrieving the chaotic attractors. The machine is trained with the parameters $(\eta,\delta\beta)=(0.2,0.4)$, by which the attractors can be accurately retrieved [see Fig.~\ref{fig1}(b)]. (a) The Lorenz attractor. (b) The R\"{o}ssler attractor. (c) The Chen attractor. (d) The Chua attractor. Upper row: the corresponding binary correlation matrices. Black points represent functional links. Low row: the corresponding functional networks. Units are colored according to the community partitions. The modularities of the functional networks associated with the Lorenz, R\"{o}ssler, Chen, and Chua attractors are about $0.56$, $0.65$, $0.60$, and $0.55$, respectively.}
\vspace{-0.3cm}
\label{fig3}
\end{center}
\end{figure*}

We adopt the method introduced in Ref.~\cite{Arenas2006} to construct the functional network. To obtain the functional network representing attractor $i$, we operate the machine in the retrieval phase by inputting $\beta_i$ in the label channel. The pairwise correlation between units $i$ and $j$ in the reservoir is defined as
\begin{equation}
\rho_{ij}(T) = \langle \cos[r_i(T)-r_j(T)] \rangle,
\label{cor}
\end{equation}
which is calculated at time $T$ of the reservoir evolution. In Eq.~(\ref{cor}), the angle bracket denotes that the result is averaged over $N_a$ reservoir realizations with different random initial conditions. Given that generalized synchronization can be established between the input signals (i.e., $\beta_i$) and the reservoir, the correlation coefficients will be stabilized to steady values after the transient ($T\gg 1$). A functional link is established between units $i$ and $j$ if $\rho_{ij}$ is larger than a predefined threshold $\rho_c$. We thus obtain a binary correlation matrix $\bf{B}$, with $b_{ij}=b_{ji}=1$ if $\rho_{ij}>\rho_c$ and $b_{ij}=0$ otherwise. Finally, the functional network can be obtained by treating $\bf{B}$ as the adjacency matrix. Shown in Fig.~\ref{fig3} are the correlation matrices (the upper row) and the corresponding functional networks (the lower row) obtained in retrieving the chaotic attractors, in which the machine are trained by data processed by the parameters $(\eta,\delta\beta)=(0.2,0.4)$ [the case where the chaotic attractors are accurately retrieved; see Fig.~\ref{fig1}(b)]. The functional networks are constructed by the parameters $(T,N_a,\rho_c)=(100,1\times 10^3,0.98)$. (The results to be reported are qualitatively the same when these parameters are slightly changed). Regarding the functional networks, the questions we are interested in can be re-expressed as: (1) What are the properties of the functional networks underpinning the successful retrievals? (2) How the properties are changed with the distinguishing parameters $\eta$ and $\delta\beta$?

A distinct feature of the functional networks shown in Fig.~\ref{fig3}(b) is the presence of community structures~\cite{CN:Newman}. Specifically, the network nodes are partitioned into different clusters, with nodes within the same cluster being densely connected while the connections between nodes belonging to different clusters are sparse. Numerical analysis based on the methods introduced in Refs.~\cite{Modularity:VDB,Modularity:RL} shows that the modularities of the functional networks associated with the Lorenz, R\"{o}ssler, Chen and Chua attractors are approximately $0.56$, $0.65$, $0.60$ and $0.55$, respectively. Since units inside the reservoir are connected by random (represented by a sparse random Erd\"{o}s-R\'{e}nyi matrix), community structures are absent in the physical network. The emergence of the community structures, therefore, is due to the self-organization of the reservoir units in response to the input signals.

Another feature of the functional networks shown in Fig.~\ref{fig3}(b) is that their community structures are different from each other. For instance, the functional network associated with the Lorenz attractor [see Fig.~\ref{fig3}(b1)] contains $4$ communities, while the functional network associated with the R\"{o}ssler attractor consists of $5$ communities [see Fig.~\ref{fig3}(b2)]. As our studies focus on the working mechanism of multi-functional RC (i.e., how different attractors are learned and retrieved by a single RC), we are particularly interested in the distinguishability of the functional networks. The metric we adopt to quantify the network distinguishability is the normalized mutual information (NMI), which is defined as~\cite{Newman2004,Karrer2008,Meil2007}
\begin{equation}
I_{ij}(\mathbf{C}_i,\mathbf{C}_j)=\frac{2I(\mathbf{C}_i,\mathbf{C}_j)}{H(\mathbf{C}_i)+H(\mathbf{C}_j)},
\end{equation}
where $\mathbf{C}_j$ and $\mathbf{C}_j$ are the community partitions (assignments) of the functional networks associated with attractors $i$ and $j$, respectively, $H(\mathbf{C})$ is the information entropy, and $I(\mathbf{C}_i,\mathbf{C}_j)$ is the mutual information between $\mathbf{C}_i$ and $\mathbf{C}_j$. (Please see Refs.~\cite{Newman2004,Karrer2008,Meil2007} for more details about the definitions.) Briefly, NMI measures how much information we can learn about $\mathbf{C}_i$ by knowing $\mathbf{C}_j$, i.e., the similarity between the two community partitions. We have $I_{ij}\in [0,1]$, and the larger (smaller) is $I_{ij}$, the more similar (different) is the community partitions $\mathbf{C}_j$ and $\mathbf{C}_j$. For the functional networks shown in Fig.~\ref{fig3}(b), the pairwise NMIs calculated from the corresponding community partitions are listed in Tab.~\ref{tab1}. We see that the values of $I_{ij}$ are attractor-dependent but are all below $0.5$ [two networks are regarded as clearly distinguishable if $I<0.3$~\cite{Newman2004,Karrer2008,Meil2007}], indicating the distinct difference between the community partitions. To evaluate the overall difference of the functional networks, we calculate the averaged NMI
\begin{equation}
\left<I\right>=\frac{2}{m(m-1)}\sum_{j>i}I_{ij},
\end{equation}
with $i,j=1,\ldots,m$ being the attractor indices. For the NMIs listed in Tab.~\ref{tab1}, we have $\left<I\right>\approx 0.121$. The averaged NMI is the metric we utilize to quantify the distinguishability of the functional networks emerged in the reservoir network in retrieving the attractors.

\begin{table}[tbp]
    \caption{For the functional networks shown in Fig.~\ref{fig3}(b), the pairwise NMIs of the corresponding community partitions.}
    \vspace{0.2cm}
   \begin{ruledtabular}
    \begin{tabular}{lcccc}
        \textbf{Attractor} & \textbf{Lorenz} & \textbf{Rössler} & \textbf{Chen} & \textbf{Chua} \\
        \hline
        \textbf{Lorenz}  & N/A& 0.024 & 0.108 & 0.11 \\
        \textbf{Rössler} & 0.024 & N/A & 0.197 & 0.318 \\
        \textbf{Chen}    & 0.108 & 0.197 & N/A & 0.332 \\
        \textbf{Chua}    & 0.11 & 0.318 & 0.332 & N/A \\
    \end{tabular}
    \end{ruledtabular}
    \label{tab1}
\end{table}

\begin{figure}[tbp]
\begin{center}
\includegraphics[width=0.8\linewidth]{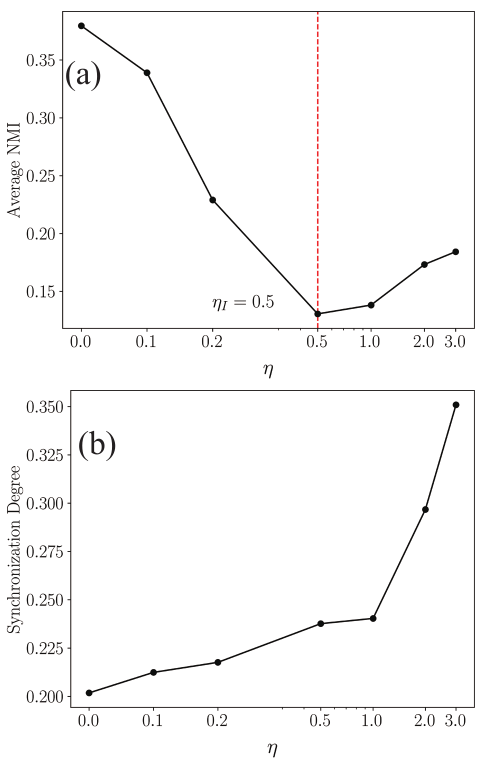}
\centering\caption{(a) The variation of the averaged NMI, $\left<I\right>$, with respect to the separation parameter, $\eta$. $\left<I\right>$ reaches its minimum at $\eta_I=0.5$. (b) The variation of the synchronization degree of the reservoir network, $\left<f\right>$, with respect to $\eta$. The machines are identical to the ones used in Fig.~\ref{fig2}(a).}
\vspace{-0.4cm}
\label{fig4}
\end{center}
\end{figure}

Is the averaged NMI a good indicator of the machine performance? To check out, we plot in Fig.~\ref{fig4}(a) the variation of $\left<I\right>$ with respect to the separation parameter, $\eta$. The set of machines is identical to the ones used in Fig.~\ref{fig2}(a), where the labeling parameter is $\delta\beta=0.4$. It is seen that by increasing $\eta$, the value of $\left<I\right>$ is first decreased and then increased, reaching its minimum at about $\eta_I=0.5$. Compared to the results shown in Fig.~\ref{fig2}(b), we see that the variation of the machine performance is well captured by that of the averaged NMI, yet with a small mismatch in the location of the optimal parameter. More specifically, the success rate reaches its maximum at $\eta_o=0.2$ in Fig.~\ref{fig2}(a), while the averaged NMI reaches its minimum at $\eta_I=0.5$ in Fig.~\ref{fig4}(a).

As $\eta$ quantifies the separation distance between the attractors in the phase space, from the perspective of memory storage and retrieval in neuropsychology~\cite{Memory:RC,Memory:RMS}, a natural expectation will be that the success rates of attractor retrievals will be increased monotonically by increasing $\eta$. However, the results in Figs.~\ref{fig2}(a) and \ref{fig4}(a) show that the distinction between the network representations of the attractors in the reservoir, namely the functional networks, is declined as $\eta$ increases from $\eta_o$ (or $\eta_I$). Why the averaged NMI is increased at large values of $\eta$? Noticing that the calculation of NMI is based on the partition of the communities while the partition is dependent of the network connectivity~\cite{Newman2004,Meil2007}, a possible explanation would be that the anomaly might be induced by the strong coherence among the reservoir units. Specifically, when the reservoir units are strongly correlated, the functional network will be dense and, consequently, the identification of the community structures becomes difficult~\cite{CN:Newman,FN:LWJ2015,CN:BS}. In particular, in the extreme case when the reservoir units are completely synchronized (i.e., the motions of the units are identical), the functional networks will be globally connected and indistinguishable from each other, resulting in $\left<I\right>=1$. A simple approach to quantifying the coherence degree of the reservoir is calculating the connecting density of the functional network, $f=2L/(N^2-N)$, with $L$ the number of functional links and $N$ the reservoir size. Shown in Fig.~\ref{fig4}(b) is the variation of the averaged connecting density (synchronization degree), $\left<f\right>=\sum f_i/m$, with respect to $\eta$. Here, $f_i$ is the connecting density of the functional network associated with the $i$th attractor. We see that $\left<f\right>$ is increased with $\eta$ monotonically.

The results presented in Fig.~\ref{fig4}(b) can be explained by the mechanism of driven synchronization, as follows. Regarding the reservoir network as a high-dimensional nonlinear dynamical system and the inputs as the externally added driving signals, the dynamics of the reservoir network will be confined into a low-dimensional space in the presence of the driving signals. Generally, the stronger the driving signals, the higher the confinement of the network dynamics. For instance, when an ensemble of chaotic oscillators is subjected to a common noise, the oscillators can be completely synchronized if the amplitude of the noise is large enough, even though the oscillators are isolated from each other~\cite{NIS:MA1994,NIS:ZCS2002}. As the separation parameter characterizes the mean values of the input signals, its increase, therefore, prompts the synchronization propensity of the reservoir network.

It is worth noting that the implications of synchronization to RC are twofold. On the one hand, the ``echo" property of RC requires that the machine outputs be independent of the initial conditions of the reservoir network~\cite{RC:Jaeger,RC:lukosevicius2009,RC:Tanaka2019}. From the viewpoint of generalized synchronization in nonlinear science~\cite{GS:Rulkov1995}, this means that the instant state of the reservoir is correlated to that of the target system by a complicated function~\cite{RC:LZ2018,RC:LZX2020}. The state of generalized synchronization is stable only when the strength of the interaction between the driving system (which generates the inputs) and the reservoir is larger than a critical value~\cite{GS:Rulkov1995}. In this regard, the increase of the separation parameter enhances the synchronization and the ``echo" property, which is beneficial for machine learning. However, if the separation parameter is too large and the interaction is too strong, the dynamics of the reservoir will be confined to a very low-dimensional space. In this case, the synchronization state is very stable, but the synchronization manifold is highly simplified~\cite{RC:LZ2018,RC:LZX2020}. [For example, when the reservoir units are completely synchronized, the synchronization manifold has the same dimension as a single unit. In such an extreme case, the reservoir behaves as consisting of a single unit (i.e., the effective size of the reservoir equals one) and is unable to encode the input signals.] Hence, by increasing the separation parameter, the complexity of the reservoir dynamics will be reduced (i.e., the number of positive conditional Lyapunov exponents is decreased~\cite{RC:LZ2018}), resulting in the decreased learning capacity. The competition between the stability and complexity of the synchronization dynamics gives rise to the optimal separation parameter by which the machine performance is maximized. This mechanism is general for mono-functional RCs designed for learning a single attractor. For multi-functional RC, the performance of the machine (i.e., the success rates for retrieving the attractors) is not only affected by the synchronization degree of the reservoir network, but also dependent on the distinguishability of the functional networks. This explains why the averaged NMI reaches its minimum at $\eta_I=0.5$ in Fig.~\ref{fig4}(a) while the success rate is maximized at $\eta_o=0.2$ in Fig.~\ref{fig2}(a).

Combining the results of functional networks and reservoir synchronization, we now are able to give a global picture on the impact of the separation parameter on the performance of multi-functional RC. When the separation parameter is small, whereas the dynamics of the reservoir is complicated enough for representing the attractors, the functional networks are unstable (as synchronization is not established between the inputs and the reservoir) and indistinguishable (the averaged NMI is large), resulting in the poor retrieval performance as depicted in Fig.~\ref{fig1}(c) under the parameter $\eta=0$. When the separation parameter is medium, the functional network is stable and distinguishable, and the dynamics of the reservoir network is still complex enough. In this case, the attractors can be successfully stored and retrieved, as depicted in Fig.~\ref{fig1}(b) under the parameter $\eta=0.2$. When the separation parameter is large [e.g., the case of $\eta=3$ shown in Fig.~\ref{fig1}(d)], though the functional networks are stable, the dynamics of the reservoir is oversimplified and is unable to represent the attractors.

\begin{figure}[tbp]
\begin{center}
\includegraphics[width=0.8\linewidth]{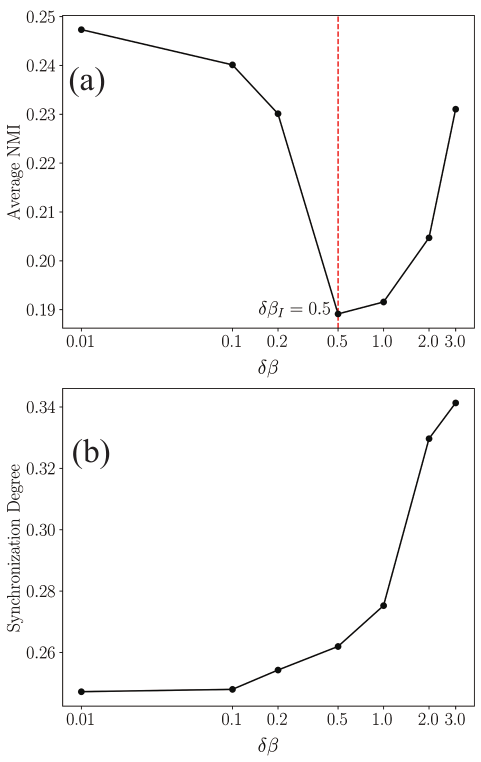}
\centering\caption{(a) The variation of the averaged NMI, $\left<I\right>$, with respect to the labeling parameter, $\delta\beta$. $\left<I\right>$ reaches its mimumum at $\delta\beta_I=0.5$. (b) The variation of the synchronization degree of the reservoir network, $\left<f\right>$, with respect to $\delta\beta$. The machines are identical to the ones used in Fig.~\ref{fig2}(b).}
\vspace{-0.4cm}
\label{fig5}
\end{center}
\end{figure}

The above picture also depicts the impact of the labeling parameter on machine performance [see Fig.~\ref{fig2}(b)]. Shown in Fig.~\ref{fig5}(a) is the variation of the averaged NMI, $\left<I\right>$, with respect to the labeling parameter, $\delta\beta$, in which the separation parameter is fixed as $\eta=0.2$. The machines are identical to the ones used in Fig.~\ref{fig2}(b). Similar to the results plotted in Fig.~\ref{fig4}(a), we see that with the increase of $\delta\beta$, the value of $\left<I\right>$ is first decreased and then increased, reaching its minimum at $\delta\beta_I=0.5$. Shown in Fig.~\ref{fig5}(b) is the variation of the averaged connecting density (synchronization degree), $\left<f\right>$, with respect to $\delta\beta$. Similar to the results shown in Fig.~\ref{fig4}(b), we see that $\left<f\right>$ is increased monotonically with $\delta\beta$. The similarity between the effects of the separation and labeling parameters is understandable, as both parameters characterize the intensity of the input signals (the former for the state channels and the latter for the label channel).

\section{Discussions and conclusion}

What happens if the label is not included in the training set? This question is of both theoretical and practical significance, as (1) the label might serve as a bifurcation parameter of the reservoir dynamics, and (2) small perturbations (parameter mismatches) are unavoidable in hardware implementations. The fact that the dynamics of the reservoir can be modified by tuning the input parameter of the control channel has been well demonstrated in the literature~\cite{RC:CK2020,KLW:2021,RC:Kim2021,RC:FHW2021,RC:multistability2022,RC:LHB2024}. For instance, in the study of PARC, it has been shown that by tunning the control parameter, the machine trained by the time series of a handful of states is able to infer the whole bifurcation diagram of the target system with a high precision~\cite{KLW:2021,RC:LHB2024}. The existing studies, however, are focusing on mono-functional machines. That is, the training and predicted states are governed by the same set of dynamical equations. As multi-functional RC is designed for learning multiple attractors governed by entirely different dynamics, it is intriguing to see what the retrieved attractor looks like if inputting a new label that is not included in the training set. In addition, in hardware implementations of multi-function RC, it is impossible to keep the label at precisely the desired value, rendering the robustness of retrieved attractors to label perturbations an important concern.

\begin{figure}[tbp]
\begin{center}
\includegraphics[width=1.0\linewidth]{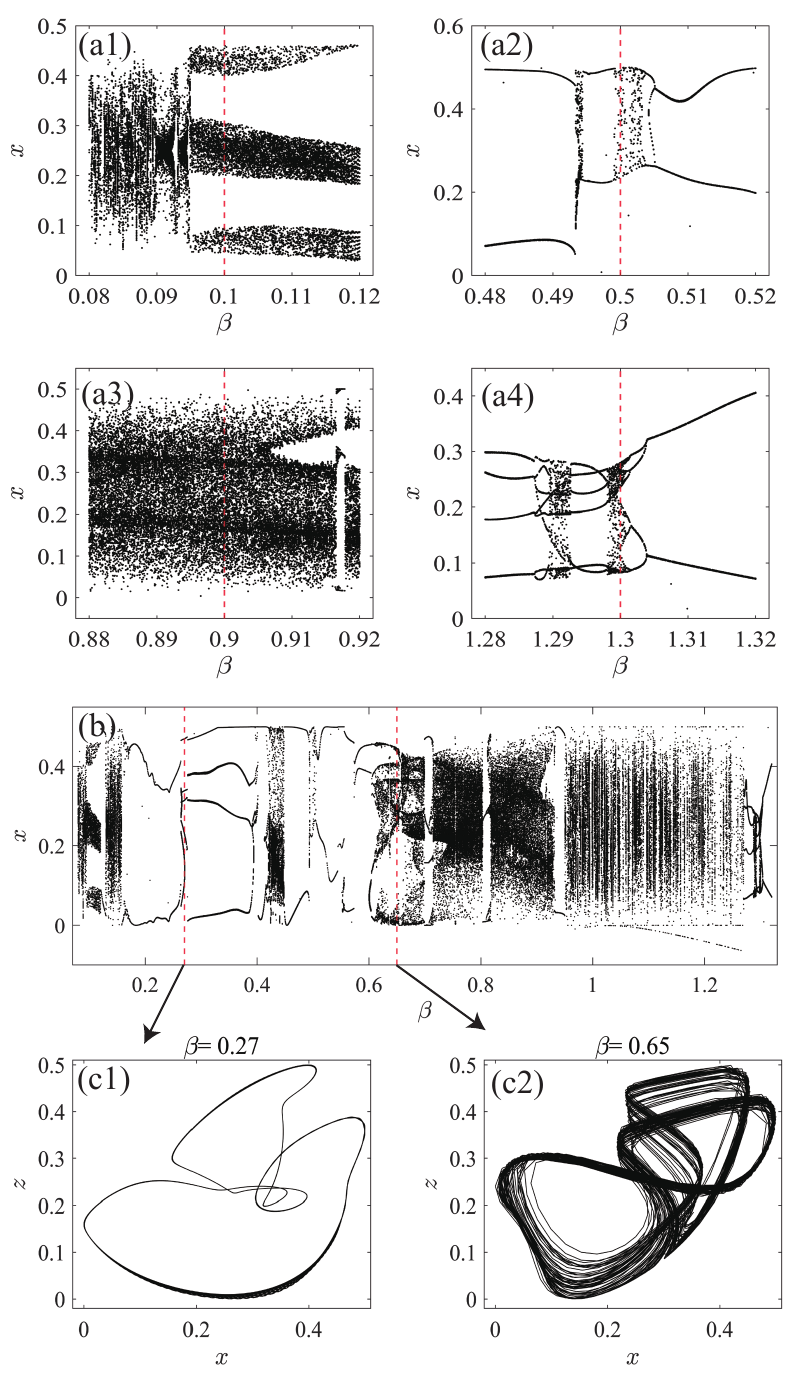}
\centering\caption{(a) Bifurcation of the retrieved attractor around the training labels. The machine is identical to the one used in Fig.~\ref{fig1}(b). Vertical lines denote the training labels. (b) Transition of the retrieved attractor with respect to the attractor label over a wide range. (c) Examples of new attractors observed in the transition. See the context for more details.}
\vspace{-0.5cm}
\label{fig6}
\end{center}
\end{figure}

By the machine used in Fig.~\ref{fig1}(b) (by which the attractors can be successfully retrieved), we tune the input label, $\beta$, around the sampling parameters [$\beta=(0.1,0.5,0.9.1.3)$], and plot in Fig.~\ref{fig6}(a) the bifurcation of the retrieved attractor. Here, the retrieved attractors are depicted by the technique of Poincare surface of section, which is $z=0.4$ for the Lorenz and Chen attractors, $z=-0.4$ for the R\"ossler attractor and $z=0$ for the Chua attractor. We see that the retrieved attractor undergoes rich bifurcations as $\beta$ changes. We also see that compared to the other three attractors, the Chen attractor [see Figure~\ref{fig6}(a3)] is more robust to label perturbations. To have a global picture on the impact of $\beta$ on the retrieved attractor, we plot in Fig.~\ref{fig6}(b) the bifurcation of the retrieved attractor across the range $\beta\in (0.08,1.32)$, which covers all the sampling labels in the training set. In plotting Fig.~\ref{fig6}(b), the centers of the retrieved attractors are shifted back to the origin (in order to better present the bifurcation diagram), and the Poincare surface is chosen as $z=0$. We see that as $\beta$ increases, the retrieved attractor is transformed among the training attractors in a complicated fashion. In particular, new types of attractors are observed in the transition regimes, as illustrated in Fig.~\ref{fig6}(c). The bifurcation diagrams highlight the complex dynamics of the reservoir network in response to the input label, and indicate also the capability of multi-functional RC in generating new chaotic dynamics~\cite{Chaos:TK2021,Chaos:CT2022}.

How about attractors of different dimensions? For convenience, the chaotic attractors we have adopted are of the same dimension. It remains a challenge if the attractors are of different dimensions. To make the machine capable of learning multiple attractors of different dimensions, the difficulty is how to coordinate the different dimensions of the input vectors. We next demonstrate that this issue can be addressed by introducing the time-delayed signals in the input layer. Let $\{n_i\}$ (with $i=1,\ldots,m$) be the dimensions of the attractors and assume that $n_i\neq n_j$, we set the dimension of the input vector, ${\bm U}(t)$, uniformly as $D_{in}=\operatorname{lcm}(n_1,\ldots,n_m)$ for all the attractors, with $\operatorname{lcm}$ the function of least common multiple. Specifically, for the state vector of the $i$th attractor, ${\bm u_i}(t)$, we expand its dimension to $D_{in}$ by introducing the new variables $[{\bm u_i}(t-\Delta t),{\bm u_i}(t-2\Delta t),\ldots,{\bm u_i}(t-\tau'\Delta t)]$, with ${\bm u_i}(t-l\Delta t)$ the state vector $l$ time steps ahead of ${\bm u_i}(t)$ and $\tau'=D_{in}/n_i$. The time series of the expanded vectors acquired from multiple attractors are then concatenated to generate the training data, which, together with the labels, are inputted into the reservoir for training the output matrix. Except for the input layer, the other settings of the machine are identical to the one used in Fig.~\ref{fig1}, and so is the implementation of the machine.

\begin{figure}[tbp]
\begin{center}
\includegraphics[width=1.0\linewidth]{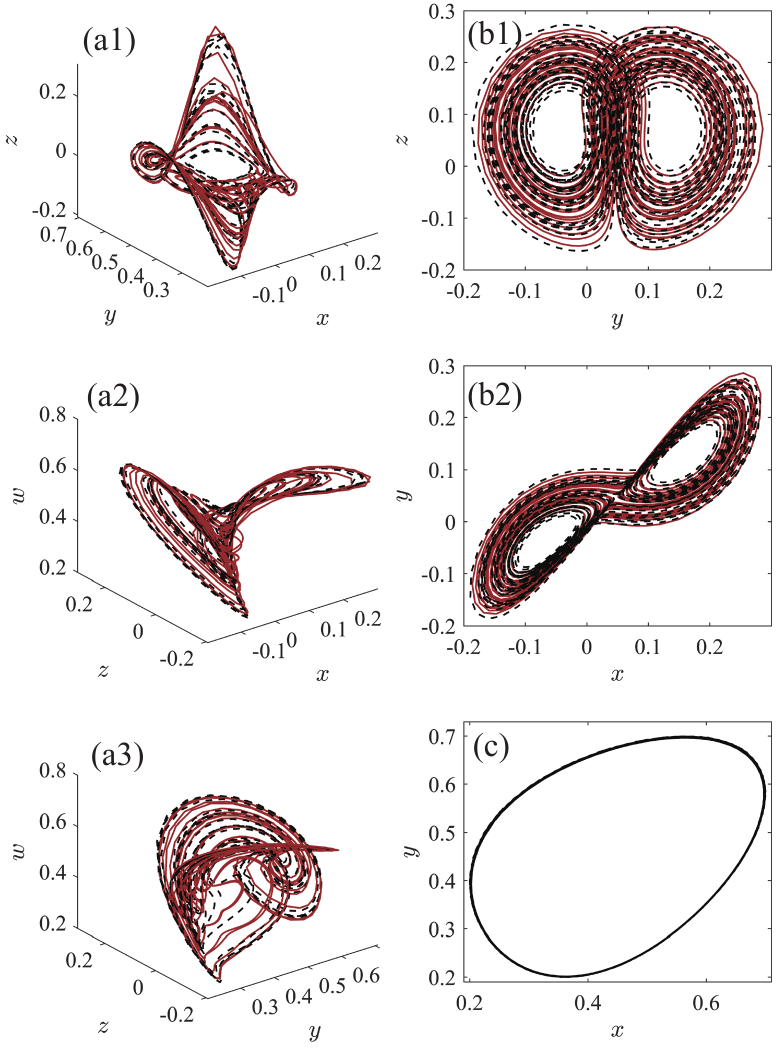}
\centering\caption{Learning multiple attractors of different dimensions. (a1-a3) The 4-dimensional hyperchaotic attractor. (b1,b2) The 3-dimensional chaotic Lorenz attractor. (c) The 2-dimensional FHN attractor. Black trajectories represent the training attractors. Red trajectories denote the retrieved attractors.}
\vspace{-0.5cm}
\label{fig7}
\end{center}
\end{figure}

We demonstrate the effectiveness of the upgraded machine by utilizing it to learn three attractors of different dimensions, including a four-dimensional hyperchaotic attractor, the three-dimensional chaotic Lorenz attractor (the one studied in Fig.~\ref{fig1}), and a two-dimensional periodic attractor generated by the FHN model. (Please see Appendix for more details about the models.) Still, the data are processed by shifting the attractors away from each other in the phase space (characterized by the separation parameter $\eta$), and the time series are labeled (by the parameter $\beta_i$). Based on the impacts of the distinguishing parameters on the machine performance revealed in the previous section (i.e., the machine performance is optimized at medium parameters), we set $\eta=0.2$ and $\delta\beta=0.5$ without optimization. Following the technique of vector expansion mentioned above, we set the dimension of the input vector as $D_{in}=\operatorname{lcm}(4,3,2)=12$. As such, the time-delay coefficients of the hyperchaotic attractor, the Lorenz attractor and the FHN attractor are, respectively, $\tau'=3$, $4$ and $6$. A new machine, with the reservoir size being still $D_{in}=100$, is then trained and optimized. The optimal hyperparameters of the new machine are $(p,\lambda,\sigma,\alpha,\xi)=(0.3,0.62,2.99,0.38,0.01)$. Shown in Fig.~\ref{fig7} are the retrieved attractors. We see that the retrieved attractors resemble the training attractors closely. Though a systematic analysis of the performance of the upgraded machine is yet to be conducted, the preliminary results do show the capability of the machine in learning multiple attractors of different dimensions.

We would like to highlight the significant implication of synchronization on the functionality of multi-functional RC and artificial neural networks in general. As a universal concept in nonlinear science, synchronization has been extensively studied by researchers from different fields over the past decades, and is recognized widely as the dynamical basis underpinning the operation of a variety of natural and man-made complex systems~\cite{Syn:Book1,Syn:Book2}. In particular, based on the phenomenon of partial synchronization, functional networks have been proposed and employed as an efficient approach to exploring the cognitive functions of the human brain~\cite{FN:BB1995,FN:KJF2011,FN:EVM2005}. As emulators of biological neural networks, artificial neural networks equipped with deep learning techniques have achieved remarkable success across numerous applications in recent years, yet their underlying working mechanisms remain elusive. Recently, efforts have been made to investigate the working mechanism of RC from the perspective of synchronization, in which some insights have been gained, e.g., the implication of generalized synchronization to machine learning~\cite{RC:LZ2018,RC:LZX2020}. The present work advances our understanding of the implication of synchronization to machine learning by showing that (1) synchronization has dual effects on machine performance, and (2) the distinguishability of the functional networks is crucial for realizing multi-functional RC. The former points out the trade-off between stability and complexity (or expressiveness) in the reservoir dynamics, while the latter emphasizes the negative effect of excessive synchronization on the learning of multiple attractors. Given the progress made in synchronization research and its pivotal role in understanding the dynamics of real-world complex systems, we believe that synchronization holds great promise for unlocking the ``black box" of artificial neural networks.

To summarize, inspired by the multifunctionality of biological neural networks, we have proposed a new scheme of reservoir computing algorithm, termed multi-functional RC, for learning multiple chaotic attractors. Two distinct features of the new machine are: (1) the attractors are of entirely different dynamics, and (2) each attractor is retrieved by inputting a specific label. The effectiveness of the machine has been demonstrated using typical chaotic attractors, and the dependence of the machine's performance on two key parameters distinguishing the attractors has been evaluated. Contrary to the intuition, it is found that the machine performance is optimized when the parameters take intermediate values. The underlying mechanism of multi-functional RC is analyzed using the method of functional networks from neuroscience, and it is found that (1) each attractor is represented by a stable, unique functional network in the reservoir, and (2) the optimal performance emerges from the balance between the stability, complexity, and distinguishability of the functional networks. Our studies introduce a new avenue for machine learning of chaotic systems, and underscore the importance of synchronization dynamics in understanding the working mechanism of artificial neural networks.

The program codes and datasets used in the current study can be obtained from Ref.~\cite{Codes}.

\begin{acknowledgments}
This work was supported by the National Natural Science Foundation of China (NNSFC) under Grant No.~12275165. XGW was also supported by the Fundamental Research Funds for the Central Universities under Grant No.~GK202202003.
\end{acknowledgments}

\appendix*
\section{Dynamical models and data preparation}

The dynamical models generating the attractors and the datasets employed in our studies are as follows. The chaotic Lorenz attractor is generated by the equations~\cite{Model:Lorenz}
\begin{equation}
\begin{cases}
\dot{x}=10(y-x),\\
\dot{y}=-xz+28z-y,\\
\dot{z}=-xy-8z/3.
\end{cases}
\label{Lorenz}
\end{equation}
In simulations, Eq.~(\ref{Lorenz}) is solved by the $4$th-order Runge-Kutta algoritm with the time step $\delta t=1\times 10^{-3}$. The attractor is chaotic, with the largest Lyapunov exponent being $\Lambda_1\approx 0.88$. The system is started from the random initial conditions, and a transient of $T=100$ is discarded before acquiring the data (time series of the system state). The data are normalized (with amplitude being $0.5$) and sampled by the time interval $\tilde\tau=0.02$. The attractor is shifted globally in the phase space by replacing $(x,y,z)$ with $(x+\eta,y+\eta,z+\eta)$, with $\eta$ being the separation parameter. A total number of $\hat{T}=9\times 10^3$ data points are acquired, which are divided into two segments. The first segment contains $T=3\times 10^3$ points, which are used as the training data. The second segment contains $T'=6\times 10^3$ points, which are used as the validating data for optimizing the machine hyperparameters.

The dynamics of the chaotic R\"{o}ssler attractor is governed by the equations~\cite{Model:Rossler}
\begin{equation}
\begin{cases}
\dot{x}=-5y-5z,\\
\dot{y}=5x+5y/2,\\
\dot{z}=10+5z(x-4).
\end{cases}
\label{Rossler}
\end{equation}
Still, the system is started from the random initial conditions, and Eq.~(\ref{Rossler}) is solved numerically by the $4$th-order Runge-Kutta algorithm with the time step $\delta t=1\times 10^{-3}$. The largest Lyapunov exponent of the attractor is $\Lambda_1\approx 0.61$. The other settings are identical to that of chaotic Lorenz attractor, including the data normalization, the time interval ($\tilde\tau$), and the lengths of the training and validating time series ($T$ and $T'$). The data points are shifted globally in the phase space by replacing $(x,y,z)$ with $(x-\eta,y-\eta,z-\eta)$.

The dynamics of the chaotic Chen attractor is governed by the equations~\cite{Model:Chen}
\begin{equation}
\begin{cases}
\dot{x}=40(y-x),\\
\dot{y}=-12x-xz+28y,\\
\dot{z}=xy-3z.
\end{cases}
\label{rosmodel}
\end{equation}
The time step used in simulating the system dynamics is $\delta t=1\times 10^{-3}$. The other settings are identical to the chaotic Lorenz oscillator. The oscillator shows chaotic motion, with the largest Lyapuono exponent being about $\Lambda_1=2.32$. The chaotic Chen attractor is shifted globally in the phase space by replacing $(x,y,z)$ with $(x-\eta,y+\eta,z+\eta)$.

The dynamics of the chaotic Chua circuit is governed by the equations~\cite{Model:Chua}
\begin{equation}
\begin{cases}
\dot{x}=c_1[7-x-g(x)],\\
\dot{y}=c_2(x-y+z),\\
\dot{z}=-c_3y,
\end{cases}
\label{chua}
\end{equation}
with $g(x)=m_1x+(m_0-m_1)(|x+1|-x-1|)/2$ the piecewise-linear function. The time step adopted for simulating the system dynamics is $\delta t=1\times 10^{-3}$. The system parameters are chosen as $(c_1,c_2,c_3,m_0,m_1)=(15.6,1,33,-8/7,-5/7)$, by which the system presents chaotic motion, with the largest Lyapunov exponent being about $\Lambda_1=0.92$. The chaotic Chua attractor is shifted in the phase space by replacing $(x,y,z)$ with $(x+\eta,y-\eta,z-\eta)$. The other settings are the same as the chaotic Lorenz oscillator.

The hyperchaotic oscillator adopted in our studies is described by the equations~\cite{Model:HypChaos2021}
\begin{equation}
\begin{cases}
\dot{x}=-y-z+w,\\
\dot{y}=x+0.1y-z,\\
\dot{z}=0.1+z(x-14)-20xw,\\
\dot{w}=-xy-0.28w.
\end{cases}
\label{hyperchaos}
\end{equation}
The time step used in simulating the system dynamics is $\delta t=1\times 10^{-3}$. The oscillator has two positive Lyapunov exponents, $\Lambda_1=0.55$ and $\Lambda_2=0.41$. The attractor is shifted in the phase space by replacing $(x,y,z,w)$ with $(x+\eta,y-\eta,z+\eta,w-\eta)$. The other settings are the same as the chaotic Lorenz oscillator.

The dynamics of the FitzHugh–Nagumo (FHN) oscillator is governed by the equations~\cite{Model:FHN}
\begin{equation}
\begin{cases}
\dot{x}=x-x^3/3-y+I_{ext},\\
\dot{y}=x+a-by.\\
\end{cases}
\label{fhn}
\end{equation}
The time step used in simulating the system dynamics is $\delta t=1\times 10^{-3}$. The parameters of the FHN oscillator are chosen as $(I_{ext},a,b)=(0.5,0.7,0.4)$, by which the system dynamics is periodic and the attractor is a two-dimensional limit cycle. The limit cycle is embedded in the three-dimensional phase space by replacing $(x,y)$ with $(x+\eta,y+\eta,0)$. The other settings are the same as the chaotic Lorenz oscillator.


\begin{thebibliography}{99}
%semiral RC works
\bibitem{RC:Maass2002} W. Maass, T. Natschlager, and H. Markram, Real-time computing without stable states: A new framework for neural computation based on perturbations, Neural Comput. {\bf 14}, 2531 (2002).

\bibitem{RC:Jaeger} H. Jaeger and H. Haas, Harnessing nonlinearity: Predicting chaotic systems and saving energy in wireless communication, Science {\bf 304}, 78 (2004).

\bibitem{RC:Pathak2017} J. Pathak, Z. Lu, B. Hunt, M. Girvan, and E. Ott, Using machine learning to replicate chaotic attractors and calculate Lyapunov exponents from data, Chaos {\bf 27}, 121102 (2017).

\bibitem{RC:LZ2018} Z. Lu, B. R. Hunt, and E. Ott, Attractor reconstruction by machine learning, Chaos {\bf 28}, 061104 (2018).

\bibitem{RC:Fan} H. Fan, J. Jiang, C. Zhang, X. G. Wang, and Y.-C. Lai, Long-term prediction of chaotic systems with machine learning, Phys. Rev. Res. {\bf 2}, 012080(R) (2020).

\bibitem{RC:digtwin} L.-W. Kong, Y. Weng, B. Glaz, M. Haile, and Y.-C. Lai, Reservoir computing as digital twins for nonlinear dynamical systems, Chaos {\bf 33}, 033111 (2023). 

\bibitem{RC:DY2024} Y. Du, Q. Li, H. Fan, M. Zhan, J. Xiao, and X. G. Wang, Inferring attracting basins of power system with machine learning, Phys. Rev. Res. {\bf 6}, 013181 (2024).

% review paper (perspectives)
\bibitem{RC:adaptiveRCLYC} S. Panahi and Y.-C. Lai, Adaptable reservoir computing: A paradigm for model-free data-driven prediction of critical transitions in nonlinear dynamical systems, Chaos {\bf 34}, 051501 (2024).

% review paper
\bibitem{RC:lukosevicius2009} M. Lukosevicius and H. Jaeger, Reservoir computing approaches to recurrent neural network training, Comput. Sci. Rev. {\bf 3}, 127 (2009).

% review paper
\bibitem{RC:Tanaka2019} G. Tanaka, T. Yamane, J. B. Heroux, R. Nakane, N. Kanazawa, S. Takeda, H. Numata, D. Nakano, and A. Hirose, Recent advances in physical reservoir computing: A review, Neural Networks {\bf 115}, 100 (2019).

%high-dimensional systems
\bibitem{RC:Pathak2018} J. Pathak, B. Hunt, M. Girvan, Z. Lu, and E. Ott, Model-free prediction of large spatiotemporally chaotic systems from data: A reservoir computing approach, Phys. Rev. Lett. {\bf 120}, 024102 (2018).

\bibitem{RC:Parlitz2018} R. S. Zimmermann, and U. Parlitz, Observing spatio-temporal dynamics of excitable media using reservoir computing, Chaos {\bf 28}, 043118 (2018).

\bibitem{RC:ParallelMachinePRL2022} K. Srinivasan, N. Coble, J. Hamlin, T. Antonsen, E. Ott, and M. Girvan, Parallel Machine Learning for Forecasting the Dynamics of Complex Networks, Phys. Rev. Lett. {\bf 128}, 164101 (2022).

\bibitem{RC:AtmosphereForecastOtt2020} I. Szunyogh, T. Arcomano, J. Pathak, A. Wikner, B. Hunt, and E. Ott, A Machine Learning-Based Global Atmospheric Forecast Model, Geophys. Res. Lett. {\bf 47}, e2020GL087776 (2020).

%transfer learning
\bibitem{RC:CK2020} C. Klos, Y. F. K. Kossio, S. Goedeke, A. Gilra, and R.-M. Memmesheimer, Dynamical learning of dynamics, Phys. Rev. Lett. {\bf 125}, 088103 (2020).

\bibitem{RC:Guo2021} Y. L. Guo, H. Zhang, L. Wang, H. W. Fan, J. H. Xiao, and X. G. Wang, Transfer learning of chaotic systems, Chaos {\bf 31}, 011104 (2021).

\bibitem{KLW:2021} L.-W. Kong, H.-W. Fan, C. Grebogi, and Y.-C. Lai, Machine learning prediction of critical transition and system collapse, Phys. Rev. Res. \textbf{3}, 013090 (2021).

\bibitem{RC:Kim2021} J. Z. Kim, Z. Lu, E. Nozari, G. J. Pappas, and D. S. Bassett, Teaching recurrent neural networks to infer global temporal structure from local examples, Nat. Mach. Intell. {\bf 3}, 316 (2021).

\bibitem{RC:FHW2021} H. Fan, L.-W. Kong, Y.-C. Lai, and X. G. Wang, Anticipating synchronization with machine learning, Phys. Rev. Res. {\bf 3}, 023237 (2021).

\bibitem{RC:multistability2022} M. Roy, S. Mandal, C. Hens, A. Prasad, N.V. Kuznetsov, and M. D. Shrimali, Model-free prediction of multistability using echo state network, Chaos {\bf 32}, 101104 (2022).

\bibitem{RC:LHB2024} H. Luo, Y. Du, H. Fan, X. Wang, J. Guo, and X. G. Wang, Reconstructing bifurcation diagrams of chaotic circuits with reservoir computing, Phys. Rev. E {\bf 109}, 024210 (2024).

\bibitem{GenerativeLearning2024} W. Gilpin, Generative learning for nonlinear dynamics, Nature Reviews Physics {\bf 6}, 194 (2024).

%multiple learning

\bibitem{RC:LZX2020} Z. Lu and D. S. Bassett, Invertible generalized synchronization: A putative mechanism for implicit learning in neural systems, Chaos {\bf 30}, 063133 (2020).

\bibitem{RC:Flynn2021} A. Flynn, V. A. Tsachouridis, and A. Amann, Multifunctionality in a reservoir computer, Chaos {\bf 31}, 013125 (2021).

%multifunctionality
\bibitem{MF:PAG} P. A. Getting, Emerging principles governing the operation of neural networks, Annu. Rev. Neurosci. {\bf 12}, 185 (1989).

\bibitem{MF:KLB} K. L. Briggman and W. B. Kristan, Multifunctional pattern-generating circuits, Annu. Rev. Neurosci. {\bf 31}, 271 (2008).

%multistability
\bibitem{MF:NP} N. Pisarchik and U. Feudel, Control of multistability, Phys. Rep. {\bf 540}, 167 (2014).

%brain memory 
\bibitem{CAM:1992} L. Chisvin and R. J. Duckworth, Content-addressable and associative memory, Advances in Computers {\bf 34}, 159 (1992).

\bibitem{Memory:RC} R. Chaudhuri and I. Fiete, Computational principles of memory, Nat. Neurosci. {\bf 19}, 394 (2016).

\bibitem{Memory:RMS} R. M. Shiffrin and R. C. Atkinson, Storage and retrieval processes in long-term memory, Psychol. Rev. {\bf 76}, 179 (1969).

\bibitem{Memory:LN} L. Nyberg, R. Habib, A. R. McIntosh, and E. Tulving, Reactivation of encoding-related brain activity during memory retrieval, Proc. Natl. Acad. Sci. U.S.A. {\bf 97}, 11120 (2000). 
%https://doi.org/10.1073/pnas.97.20.11120

\bibitem{Memory:PK} P. Khader, M. Burke, S. Bien, C. Ranganath, and F. R\"{o}sler, Content-specific activation during associative long-term memory retrieval, NeuroImage {\bf 27}, 805 (2005). 
%https://doi.org/10.1016/j.neuroimage.2005.05.006

\bibitem{Memory:JEK} J. E. Kragel, {\it et al.}, Similar patterns of neural activity predict memory function during encoding and retrieval, NeuroImage {\bf 155}, 60 (2017). 
%https://doi.org/10.1016/j.neuroimage.2017.03.042

%!TEX encoding = UTF-8 Unicodefunctional networks
\bibitem{FN:CR2005} C. Ranganath, A. Heller, M. X. Cohen, C. J. Brozinsky, and J. Rissman, Functional connectivity with the hippocampus during successful memory formation, Hippocampus {\bf 15}, 997 (2005). %https://doi.org/10.1002/hipo.20141

\bibitem{FN:EB2009} E. Bullmore and O. Sporns, Complex brain networks: graph theoretical analysis of structural and functional systems, Nat. Rev. Neurosci. {\bf 10}, 186 (2009).

\bibitem{FN:HSY2023} S. Y. Huo and Z. H. Liu, Condensation of eigenmodes in functional brain network and its correlation to chimera state, Communications Physics {\bf 6}, 285 (2023).

\bibitem{KLW2024} L.-W. Kong, G. A. Brewer, and Y.-C. Lai, Reservoircomputing based associative memory and itinerancy for complex dynamical attractors, Nat. Commun. {\bf 15}, 4840 (2024).

%DV
\bibitem{Noise:SR2023} Z.-M. Zhai, L.-W. Kong and Y.-C. Lai, Emergence of a stochastic resonance in machine learning, Phys. Rev. Res. {\bf 5}, 033127 (2023).

\bibitem{Optimization:LY} L. Yang and A. Shami, On hyperparameter optimization of machine learning algorithms: Theory and practice, Neurocomputing {\bf 415}, 295 (2020).

%%functional network, method and metrics

\bibitem{FN:ZCS2006} C. S. Zhou, L. Zemanova, G. Zamora, C. C. Hilgetag, and J. Kurths, Hierarchical organization unveiled by functional connectivity in complex brain networks, Phys. Rev. Lett. {\bf 97}, 238103 (2006).

\bibitem{FN:LMH2010} M. Li, X. G. Wang, and C. H. Lai, Evolution of functional subnetworks in complex systems, Chaos {\bf 20}, 045114 (2010).

\bibitem{FN:LWJ2015} W. Lin, Y. Wang, H. Ying, Y.-C. Lai, and X. G. Wang, Consistency between functional and structural networks of coupled nonlinear oscillators, Phys. Rev. E {\bf 92}, 012912 (2015).

\bibitem{FN:BB1995} B. Biswal, F. Z.  Yetkin, V. M. Haughton, and J. S. Hyde, Functional connectivity in the motor cortex of resting human brain using echo-planar MRI, Magnetic Resonance in Medicine {\bf 34}, 537 (1995).

\bibitem{FN:KJF2011} K. J. Friston, Functional and Effective Connectivity: A Review, Brain Connect. {\bf 1}, 13 (2011).

\bibitem{FN:EVM2005} V. M. Eguiluz, D. R. Chialvo, G. A. Cecchi, M. Baliki, and A. V. Apkarian, Scale-Free Brain Functional Networks, Phys. Rev. Lett. {\bf 94}, 018102 (2005).

\bibitem{CN:Newman} M. E. J. Newman, The structure and function of complex networks, SIAM Rev. {\bf 45}, 167 (2003).

\bibitem{CN:BS} S. Boccaletti, V. Latora, Y. Moreno, M. Chavez, and D.-U. Hwang, Complex networks: Structure and dynamics, Phys. Rep. {\bf 424}, 175 (2006).

%RC relation to synchronization

\bibitem{GS:Rulkov1995} N. F. Rulkov, M. M. Sushchik, L. S. Tsimring, and H. D. Abarbanel, Generalized synchronization of chaos in directionally coupled chaotic systems, Phys. Rev. E. {\bf 51}, 980 (1995).

%\bibitem{GS:ZM2003} M. Zhan, X. G. Wang, X. Gong, G. W. Wei, and C.-H. Lai, Complete synchronization and generalized synchronization of one-way coupled time-delay systems, Phys. Rev. E. {\bf 68}, 036208 (2003).

\bibitem{RC:GSTL2019} T. Lymburn, D. M. Walker, M. Small, and T. Jungling, The reservoir's perspective on generalized synchronization, Chaos {\bf 29}, 093133 (2019).

\bibitem{RC:VerzelliChaos2021} P. Verzelli, C. Alippi, and L. Livi, Learn to synchronize, synchronize to learn, Chaos {\bf 31}, 083119 (2021).

\bibitem{RC:WL2022} L. Wang, H. Fan, J. Xiao, Y. Lan, and X. G. Wang, Criticality in reservoir computer of coupled phase oscillators, Phys. Rev. E {\bf 105}, L052201 (2022).

%FN method
\bibitem{Arenas2006} A. Arenas, A. Diaaz-Guilera, and C. J. Perez-Vicente, Synchronization reveals topological scales in complex networks, Phys. Rev. Lett. {\bf 96}, 114102 (2006).

%modularity

\bibitem{Modularity:VDB} V. D Blondel, J.-L. Guillaume, R. Lambiotte, and E.e Lefebvre, Fast unfolding of communities in large networks, J. Stat. Mech. Theory Exp. {\bf 10}, 10008 (2008).

\bibitem{Modularity:RL} R. Lambiotte, J.-C. Delvenne, and M. Barahona, Random Walks, Markov Processes and the Multiscale Modular Organization of Complex Networks, IEEE Trans. Netw. Sci. Eng {\bf 1}, 76 (2014).

%characterizing complex networks

\bibitem{Newman2004} M. E. J. Newman and M. Girvan, Finding and evaluating community structure in networks, Phys Rev E {\bf 69}, 026113 (2004).

\bibitem{Karrer2008} B. Karrer, E. Levina, and M. E. J. Newman, Robustness of community structure in networks, Phys. Rev. E {\bf 77}, 046119 (2008).

\bibitem{Meil2007} M. Meila, Comparing clusterings -- an information based distance, J. Multivar. Anal. {\bf 98}, 873 (2007).

%https://doi.org/10.1063/1.3523297

%noised-induced chaos synchronization

\bibitem{NIS:MA1994} A. Maritan and J. R. Banavar, Chaos, noise, and synchronization, Phys. Rev. Lett. {\bf 72}, 1451 (1994).

\bibitem{NIS:ZCS2002} C. Zhou and J. Kurths, Noise-Induced Phase Synchronization and Synchronization Transitions in Chaotic Oscillators, Phys. Rev. Lett. {\bf 88}, 230602 (2002).

%chaos generation

\bibitem{Chaos:TK2021} K. Tian, C. Grebogi, and H.-P. Ren, Chaos generation with impulse control: Application to non-chaotic systems and circuit design, IEEE Transactions on Circuits and Systems I: Regular Papers {\bf 68}, 3012 (2021).

\bibitem{Chaos:CT2022} T. L. Carroll, Creating new chaotic signals with reservoir computers, Chaos, Solitons \& Fractals {\bf 164}, 112688 (2022).

%synchronization reviews

\bibitem{Syn:Book1} A. S. Pikovsky, M. G. Rosenblum, and J. Kurths, {\it Synchronization: A Universal Concept in Nonlinear Science} (Cambridge University Press, Cambridge, 2001).

\bibitem{Syn:Book2} S. Strogatz, {\it Sync: The Emerging Science of Spontaneous Order} (Hyperion, New York, 2003).

%chaos models

\bibitem{Model:Lorenz} E. N. Lorenz, Deterministic nonperiodic flow, Journal of the Atmospheric Sciences {\bf 20}, 130 (1963).

\bibitem{Model:Rossler} O. E. R\"{o}ssler, An Equation for Continuous Chaos, Phys. Lett. A {\bf 57}, 397 (1976).

\bibitem{Model:Chen} G. R. Chen, Yet another chaotic attractor, International Journal of Bifurcation and Chaos {\bf 9}, 1465 (1999).

\bibitem{Model:Chua} L. O. Chua, Chua’s circuit: An overview ten years later, J. Circ. Syst. Comput. {\bf 04}, 117 (1994).

%\bibitem{Model:Hyperchaos} O. E. R\"{o}ssler, An equation for hyperchaos, Phys. Lett. {\bf 71A}, 155 (1979).

\bibitem{Model:HypChaos2021} G. Laarem, A new 4-D hyper chaotic system generated from the 3-D R\"{o}sslor chaotic system, dynamical analysis, chaos stabilization via an optimized linear feedback control, it’s fractional order model and chaos synchronization using optimized fractional order sliding mode control, Chaos, Solitons \& Fractals {\bf 152}, 111437 (2021).

\bibitem{Model:FHN} R. FitzHugh, Impulses and physiological states in theoretical models of nerve membrane, Biophysical J. {\bf 1}, 445 (1961).

\bibitem{Codes} https://github.com/Xingang-Wang/MRFC




 

\end{thebibliography}
\end{document}